\documentclass[aps,floats,showpacs,twocolumn,prl]{revtex4}
\usepackage{epsfig}
\usepackage{amssymb}
\usepackage[dvips]{color}
\usepackage[latin1]{inputenc}
\usepackage{graphicx}
\usepackage{epsfig}

\newcommand{\nc}{\newcommand}
\nc{\be}{\begin{equation}}
\nc{\ee}{\end{equation}}
\nc{\bea}{\begin{eqnarray}}
\nc{\eea}{\end{eqnarray}}
\nc{\bt}{\begin{tabular}}
\nc{\et}{\end{tabular}}
\nc{\ba}{\begin{array}}
\nc{\ea}{\end{array}}
\nc{\bvec}{\mathbf}
\nc{\vecna}{\mbox{\boldmath $\nabla$}}
\nc{\erm}{{\rm e}}
\nc{\Lc}{{\cal L}}
\nc{\um}{\frac{1}{2}}
\nc{\vbf}{\mbox{\boldmath $v$}}
\nc{\jbf}{\mbox{\boldmath $j$}}
\nc{\dis}{\displaystyle}
\nc{\rd}{\rm d}
\nc{\tcb}{\textcolor{blue}}
\nc{\tcr}{\textcolor{red}}
\nc{\Hc}{{\cal H}}

\begin{document}

\title{A generalization of the Ginzburg-Landau theory to $p$-wave superconductors}

\author{E. Di Grezia}
\email{digrezia@na.infn.it}
\affiliation{Universit\`a Statale di Bergamo, Facolt\`a di
Ingegneria, viale Marconi 5, 24044 Dalmine (BG), Italy
and Istituto Nazionale di Fisica Nucleare, Sezione di Milano, via
Celoria 16, I-20133 Milan, Italy}
\author{S. Esposito}
\email{salvatore.esposito@na.infn.it}%
\affiliation{Dipartimento di Scienze Fisiche, Universit\`{a} di
Napoli ``Federico II'' and Istituto Nazionale di Fisica Nucleare,
Sezione di Napoli, Complesso Universitario di Monte S. Angelo, via
Cinthia, I-80126 Napoli, Italy}
\author{G. Salesi}
\email{salesi@unibg.it} %
\affiliation{Universit\`a Statale di Bergamo, Facolt\`a di
Ingegneria, viale Marconi 5, 24044 Dalmine (BG), Italy
and Istituto Nazionale di Fisica Nucleare, Sezione di Milano, via
Celoria 16, I-20133 Milan, Italy }

\begin{abstract}

We succeed to build up a straightforward theoretical model for spin-triplet
$p$-wave superconductors by introducing in Ginzburg-Landau theory a second order parameter and a suitable interaction between the two mean fields.


\end{abstract}

\maketitle

\section{``Doubling'' the Ginzburg-Landau theory}

In early works based on the mean-field ``macroscopic'' approach \cite{L9} superfluidity is described in terms of a single order parameter $\phi_1\equiv\sqrt{\rho}\,{\rm e}^{i\theta}$. Subsequently Ziman \cite{Ziman} considered besides the phonon propagation also the  roton motion by introducing, in addition to the usual couple of fields $\rho$ and $\theta$, other two real fields, $\chi$ and $\psi$, entering the Clebsh term of the most general expansion of a generic velocity field:
\ $\vbf = \vecna\theta + \chi\vecna\psi\,.$ \ Correspondingly, besides the usual complex quantum field $\phi_1$, Ziman introduced a second complex quantum field accounting for rotational motions, \
$\phi_2\equiv(\psi_a+i\psi_b)/2\hbar$, \ where the real and imaginary parts are
linked to $\chi$ and $\psi$ as follows: \
$
\psi_a \equiv \sqrt{2\rho\chi}\psi\,, \ \psi_b \equiv \sqrt{2\rho\chi}\,.
$ \
Hence the current density can be re-written as a sum of a irrotational (superfluid) part and of a non-potential (non-superfluid) part \ \ $\jbf =\rho\vbf = \rho(\vecna\theta + \chi\vecna\psi) = \dis\frac{i}{2}\hbar(\phi_1^\star\vecna\phi_1 - \phi_1\vecna\phi_1^\star) + \dis\frac{i}{2}\hbar(\phi_2^\star\vecna\phi_2 - \phi_2\vecna\phi_2^\star)$. \
Finally the complete Hamiltonian, entailing also roton kinetic energy and phonon-roton interaction terms, can be written straightforwardly as the sum of the total kinetic energy and of a suitable  pressure potential \ $\Hc = \um\rho\vbf^2 + \rho\dis\int_{\rho_0}^{\rho}\frac{p-p_0}{\rho^2}\rd\rho\,.$ \
Therefore two independent complex order parameters are needed for a complete picture of the superfluid dynamics and the original Landau irrotational mean-field theory must be ``doubled'' in order to describe the rotational degrees of freedom.
Ginzburg and Landau  \cite{GL} started from the analogy with $^3$He superfluid to derive their superconductivity theory based on the existence of an underlying mean field $\phi$ in the bulk of a superconducting medium which can be interpreted as the wave-function of the Cooper pair in its center-of-mass frame. At the same time, in a quantum field framework, the order parameter can be conceived as a self-interacting Higgs field which undergoes condensation for U(1) symmetry breaking when the temperature approaches a critical temperature.
The Ginzburg-Landau (GL) theory was built up simply taking into account the electric and magnetic properties of the Cooper pairs already discovered in the theoretical framework of the fermionic superfluidity. If, by analogy with the Ziman approach, we introduce a second order parameter in GL superconductivity theory, we, of course, will not describe rotons but actually two different superconducting systems: \textit{spin-singlet two-phase superconductors} or, as we shall show in the next section, \textit{spin-triplet one-phase superconductors}.

As regards the former, in recent works \cite{TwoTC, Term2TC, Magn2TC} we considered two scalar charged fields $\phi_{\rm w}$ and $\phi_{\rm s}$ corresponding to Cooper pairs with electrons bound by a weaker or stronger attractive force, respectively. In so doing we obtained a theoretical model for superconductors endowed with two distinct superconducting phases, since the two order parameters condensate at different critical temperatures \footnote{Let us recall that, as is proved in the above-quoted papers, the introduction of additional degrees of freedom (two complex fields rather than one) \emph{does not involve new unknown physical constants} since both scalar fields are endowed with equal \emph{bare} masses and self-interaction coupling constants.}.
In \cite{Term2TC} we found some deviations in basic thermodynamical quantities with respect to the Ginzburg-Landau one-phase superconductors. In particular, in contrast to the usual case where  only one jump in specific heat takes place at the normal-superconductor transition temperature, we actually predicted an additional discontinuity for $C_V$ when passing from a superconducting phase to the other one. Furthermore, on analyzing the magnetic behavior of such systems \cite{Magn2TC} , we found some observable differences with respect to the case of conventional Ginzburg-Landau superconductors. In particular, at low temperature the London penetration length is strongly reduced and the Ginzburg-Landau parameter $\kappa$ becomes a function of temperature. By contrast, in the temperature region between the two phase transitions, $\kappa$ is constant and the system is a type-I or a type-II
superconductor depending on the ratio between the critical temperatures.
A thermodynamical and magnetic behavior qualitatively similar to the one predicted by our model has been recently observed in MgB$_2$ \footnote{Several two-band theories \cite{2Gaps} try to explain the observed behavior of the specific heat  and the peculiar temperature-dependence of the upper and lower critical fields of MgB$_2$: probably it exists a correspondence between the \textsl{two} ``classical-macroscopic'' (since represent ``collective'' wave-functions for the condensate) order parameters in GL-like approaches as the present one, and the \textsl{two}  ``quantum-microscopic'' gaps in quasi-particle energy spectra predicted for MgB$_2$ by some BCS-like theories.} (see, for example \cite{BuzeaYamashita} and Refs.\,therein).

By introducing two {\em mutually interacting} order parameters, in the present letter we will not describe two-phases superconductors but actually \textit{spinning} Cooper pairs and rotational degrees of freedom in superconductivity.
In Bardeen, Cooper, and Schrieffer (BCS) theory for conventional superconductors
the electrons are paired in a zero total angular momentum state, with zero spin and zero orbital angular momentum: \ $J = L = S = 0$. \ As a matter of fact, in BCS superconductors the $s$-wave is shown to be the minimum energy state with maximum attraction between the electrons of a Cooper pair.
Indeed, soon after the BCS theory was advanced, Kohn and Luttinger \cite{Kohn} predicted
that if the mutual interaction is \textit{repulsive} in all partial wave channels the Cooper
pairs result to be bonded by a weak residual attraction (out of the Coulomb repulsion) in higher angular momentum channels:  the so-called \textit{Kohn-Luttinger effect}. On the other hand it actually exists a $p$-wave Cooper-pairing in superfluid $^3$He (which is, as abovesaid, the liquid counterpart of GL superconductors). Actually, we can meet $p$-wave superconductivity in certain "heavy-electron" compounds
(\textit{Heavy Fermion Systems} as, e.g., UPt$_3$) and in special materials recently discovered as, e.g., Sr$_2$RuO$_4$ \cite{Maeno}, which is the only known metal oxide displaying $p$-wave superconductivity.
Let us recall that the $p$-wave Cooper pairs are always spin-triplets ($S$=1) because of Pauli's exclusion principle applied to systems composed of a pair of particles endowed with odd ($L$=1) total orbital quantum number. Taking into account this property, in the next section we shall put forward a simple GL-like model just for spin-triplet superconductors.

\section{The model}

Let us consider a physical system described by one doublet of complex scalar fields $\phi_1, \phi_2$.
We introduce the following Lagrangian density:
\begin{eqnarray}
{\cal L}&=&-\frac{1}{4}F_{\mu \nu }F^{\mu \nu }+\left( D_{\mu
}\phi_1 \right)^{\star }\left( D^{\mu }\phi_1 \right) \nonumber \\
&+& \left( D_{\mu }\phi_2 \right)^{\star }\left( D^{\mu }\phi_2
\right)-\lambda (|\phi_1 |^{2}-\frac{1}{2}\phi_{0}^{2})^2+
\nonumber \\ & - & \lambda(|\phi_2|^{2} -\frac{1}{2}\phi_{0}^{2})^2
+ V(\phi_1,\phi_2) \label{2a}
\end{eqnarray}
where the covariant derivative $D_\mu\equiv\partial_\mu+ieA_\mu$ describes the minimal electromagnetic interaction of the two scalar fields, while the first term in the Lagrangian (with $F_{\mu \nu }\equiv \partial_\mu A_\nu -\partial_\nu A_\mu$) accounts for the kinetic \textbf{energy} of the free electromagnetic field $A_\mu$.
The complete potential term describing the interaction of the two scalar fields is composed of three different terms, \ $V=V_{\phi A}+ V_{\rm self}+ V(\phi_1, \phi_2)$, \ the first two of them describing the usual interaction between the electromagnetic field and the charged scalar field (coming from the covariant derivative), and $V_{\rm self}$ rules the self interaction of the scalar fields:
$V_{\rm self}\equiv\lambda |\phi_1 |^{4}+ \mu |\phi_2|^{4}$.
For the interaction between the two scalar fields we instead adopt the following nonlinear term:
\begin{equation} \label{v12}
V(\phi_1,\phi_2)\equiv -\frac{\lambda\phi_0^4}{8}\ln^2{\frac{\phi_1^{\vphantom{\star}}}{\phi_1^\star}
\frac{\phi_2^\star}{\vphantom{\phi_2^\star}\phi_2^{\vphantom{\star}}}}.
\end{equation}

Let us study the small fluctuations of the two scalar fields around the  minimum of the energy corresponding to $\phi_1=\phi_2=\phi_0/\sqrt{2}$ by expanding both scalar fields as
follows:
\begin{eqnarray}
&&\phi_1 \equiv \frac{1}{\sqrt
2}(\phi_0+\eta_1)\,{\rm e}^{i\theta_1/\phi_0}\label{phi1}\\
&&\phi_2 \equiv \frac{1}{\sqrt
2}(\phi_0+\eta_2)\,{\rm e}^{i\theta_2/\phi_0}
\label{phi2}
\end{eqnarray}
where $\eta_1$, $\eta_2$, $\theta_1$, $\theta_2$ are real fields.
Exploiting these definitions, the above interaction term can be written more simply as follows
\be
V(\phi_1,\phi_2) = \frac{\lambda\phi_0^2}{2}(\theta_1-\theta_2)^2\,.
\ee
Notice that $V(\phi_1,\phi_2)$ is definite-positive, then describing a \textit{repulsion} between the two fields with strength $\lambda\phi_0^2$ equal to the mass squared $m_W^2$ (see below). Note also that
$V(\phi_1,\phi_2)$ corresponds to the main term of the expansion for small phase differences of the Legget interaction \cite{Legget} \ $\gamma(\phi_1^\star\phi_2 + \phi_1\phi_2^\star)$.

By inserting Eqs.\,(\ref{phi1},\ref{phi2}) into the Lagrangian density (\ref{2a}) and performing the gauge transformation:
$A_{\mu }\rightarrow A_{\mu }+\partial_{\mu
}\Lambda $ with:
\begin{equation}
\Lambda \equiv -\frac{1}{2e\phi_0}(\theta_1+\theta_2)\,, \label{5a}
\end{equation}
we obtain the following Lagrangian up to quadratic terms in the fields:
\begin{eqnarray}
{\cal L}&\simeq&-\frac{1}{4}F_{\mu \nu }F^{\mu \nu
}+e^{2}\phi_{0}^2A_\mu A^\mu\nonumber \\&+&\frac{1}{2}\partial_{\mu }\eta_1\partial^{\mu
}\eta_1+\frac{1}{2}\partial_{\mu }\eta_2\partial^{\mu }\eta_2
 \nonumber\\
&+&\frac{1}{2}\partial_{\mu}(\theta_1 -
\theta_2)\partial^{\mu}(\theta_1 -
\theta_2)\nonumber\\
&+&\lambda\phi_{0}^{2}\eta_1^2+ \lambda
\phi_{0}^{2}\eta_2^2+\frac{\lambda \phi_0^2}{2}(\theta_1 -
\theta_2)^2. \label{12a}
\end{eqnarray}

Let us set\noindent:
\begin{equation}
\eta_3\equiv\frac{1}{\sqrt{2}}(\theta_1 - \theta_2)\label{13a}
\end{equation}
and define the triplet field: $W_a\equiv(\eta_1,\eta_2,\eta_3)$.
The Lagrangian describing our physical system now becomes:
\begin{eqnarray}
{\cal L}&\simeq&-\frac{1}{4}F_{\mu \nu }F^{\mu \nu }+ m_A^{2}A_\mu
A^\mu
\nonumber \\ &+& \frac{1}{2}(\partial_{\mu }W_a)(\partial^{\mu
}W_a)+m_W^{2}W_aW_a\label{14a}
\end{eqnarray}
with
\begin{equation}
m_A^2= e^{2}\phi_{0}^2 , \qquad \quad
m_W^2=\lambda\phi_{0}^{2}.\label{15a}
\end{equation}
As a result, of the original four degrees of freedom embedded into two charged (complex) scalar fields, only one of them is disappeared giving rise to a massive photon, as in the standard GL model. However, by virtue of the interaction potential in Eq.\,(\ref{v12}), the remaining three degrees of freedom
all have the same mass, and can thus be combined to form a triplet field $W$ (i.e. a triplet spinorial representation of SU(2)), suitable to describe a $p$-wave superconductor. We stress that, notwithstanding the simultaneous condensation of two real degrees of freedom, the key point in our model is the particular interaction term we have
introduced, which prevents a gauge transformation to re-absorb one more degree of freedom (only the \textit{sum} of the phases turns out to be "eaten", but not even the \textit{difference}). Such a very peculiar interaction breaks the isotropy of the original medium and allows pairs of electrons to arrange into possible $S=1$ (instead of $S=0$) Cooper pairs. As a matter of fact, the emergence of a triplet field is a signal of the occurred ``anisotropization'' of the system, which can no more be described by a singlet scalar field.

\section{Application to triplet superconductors}

The order parameter describing $p$-wave superconductors may be associated in our model to the above triplet Higgs field $W_a$ which is responsible of the U(1) spontaneous symmetry breaking occurring during the normal state-superconducting-phase transition. Therefore, from the Lagrangian (\ref{2a}), the effective free energy density at finite temperature $T$, resulting from the quantum fields calculation, including one one-loop radiative corrections \cite{Bailin,NXC}, is giving by
\begin{eqnarray}
F(T) &=& F_{\rm n}(T) + a(T)|\phi_1|^2 +a(T)|\phi_2|^2 \nonumber \\ &+&
\lambda|\phi_1|^4 + \lambda|\phi_2|^4 +  a(T)\frac{\phi_0^2}{8}\left|\ln{\frac{\phi_1^{\vphantom{\star}}}{\phi_1^\star}
\frac{\phi_2^\star}{\vphantom{\phi_2^\star}\phi_2^{\vphantom{\star}}}}\right|^2
\end{eqnarray}
where
\be
a(T) = - m_W^2 + \frac{\lambda + e^2}{4}\,T^2\,,
\label{a_I}
\ee
label $n$ referring to the normal phase.
The parameter $a$ vanishes when the temperature approaches a critical value
given by
\be
T_c = \sqrt{\frac{4m_W^2}{\lambda + e^2}}\,.
\label{T1}
\ee
Below $T_c$ the expectation values of the scalar fields $\phi_1$ and $\phi_2$
which minimize the free energy function results to be
\be
|\phi_1(T)| = |\phi_2(T)| = \sqrt{-\frac{a(T)}{2\lambda}}\, ,
\label{etaquad}
\ee
while the third degree of freedom defined in Eq.\,(\ref{13a}) fluctuates around the zero  expectation value, corresponding to $\theta_1 = \theta_2$. This last occurrence directly
comes from the fact that the non-linear characteristic potential term in Eq. (\ref{v12}) is non-negative definite, so that the minimum of the free energy is reached when it vanishes. In this case, our model practically reduces to a ``simple'' doubling of the standard GL theory making recourse to two scalar order parameters.
As a consequence, it is very easy to re-obtain the usual main properties for $p$-wave superconductors considered here.

The London penetration length of the magnetic field inside the superconductor arises due to the presence of a massive photon, that is:
\begin{equation}
\delta = \frac{1}{m_A} = \frac{1}{e \phi_0} \, ,
\end{equation}
while the coherence length of the Cooper pairs described by the triplet scalar field is given by:
\begin{equation}
\xi = \frac{1}{m_W} = \frac{1}{\phi_0 \sqrt{\lambda}} = \frac{\xi_0}{\dis\sqrt{1- \frac{T^2}{T_c^2}}} \, .
\end{equation}
The \textit{critical magnetic field} $H_c$, measuring the condensation energy
$F(T) - F_{\rm n}(T) = - \mu_0 H_c^2 /2$ of the superconductor system can be obtained as follows:
\begin{equation}
H_c^2 = \frac{1}{\mu_0} \, \frac{a^2(T)}{\lambda} = H_{c0}^2 \left( 1- \frac{T^2}{T_c^2} \right)^2 \,.
\end{equation}
By taking the derivative of the free energy function with respect to temperature, we easily get the entropy gain with respect to the normal phase:
\begin{equation}
S-S_n = \frac{\partial \;}{\partial T} \left( - \frac{a^2(T)}{2\lambda} \right) = S_0
\left( 1- \frac{T^2}{T_c^2} \right) \frac{T}{T_c} \, .
\end{equation}
Finally, we can write down the expected discontinuity of the specific heat at the critical point:
\begin{equation}
\Delta C_V = T \, \frac{\partial \;}{\partial T} \left( S - S_n \right) = S_0 \left( 1- 3 \, \frac{T^2}{T_c^2} \right) \frac{T}{T_c} .
\end{equation}

\section{Conclusions}

We have extended the standard GL theory in order to describe $p$-wave superconductors by means of
two mutually interacting order parameters which condensate simultaneously at a same critical temperature (since the $\lambda\phi^4$ self-interaction is the same for both fields). After the condensation via Higgs mechanism, we remain with three massive degrees of freedom (besides a massive photon related to the Meissner-effect) which can be put in correspondence to the three components of a $S=1$ triplet mean-field describing spinning $p$-wave Cooper pairs.
In our model the main magnetic and thermodynamical (discontinuity in the specific heat included) properties of $p$-wave superconductors turn out to be essentially the same as for conventional $s$-wave superconductors.

\section*{Ackowledgements}

\noindent This work has been partially supported by Istituto Nazionale di Fisica
Nucleare and Ministero dell'Università e della Ricerca Scientifica.

\

\end{document}